\documentstyle [12pt,fleqn]{article}

\begin{document}


\newcommand{\beq}{\begin{equation}}
\newcommand{\eeq}{\end{equation}}
\newcommand{\bea}{\begin{eqnarray}}
\newcommand{\eea}{\end{eqnarray}}

\newcommand{\Prop}{\sim}


\begin{center}
{\Large ORDER AND CHAOS IN QUANTUM IRREGULAR }\\
{\Large SCATTERING: WIGNER'S TIME DELAY}
\\ \ \\
{Bruno Eckhardt}
\\ \ \\
{Fachbereich Physik der Philipps-Universit\"{a}t}\\
{Renthof 6, D-3550 Marburg}\\
\end{center}

\centerline{INTRODUCTION}

Nearly thirty years of chaos research have shown that chaos is
an ubiquitous phenomenon.  Almost any
dynamical system with a sufficiently large phase space described by
nonlinear equations shows some kind of complexity
in its time evolution, i.e.\ chaos. The degree of complexity is measured by
topological entropy, Lyapunov exponents, decay of correlations, and,
in the case of dissipative systems, various fractal dimensions.
Examples abound, ranging from nonlinear oscillators to hydrodynamical
systems and the motion of planets and asteroids (Schuster 1988).
Nevertheless, these investigations have also revealed significant
differences between an irregular looking chaotic signal and a
really and truly random one, due to `order within chaos'
(the title of a conference, Campbell and Rose 1983,
and a book, Berg{\'e}, Pomeau and Vidal 1988).
One particular example of this order and coherence in chaos will be
exploited below.

The fascination excerted by `quantum chaos' and the controversy
that surrounds it derives partly from the fact that according
to the above measures, there is no chaos!
For once, there is no nonlinearity in the Schr\"odinger equation.
Secondly, quantal spectra of bounded systems are discrete,
so that the time evolution of any observable
is multiply periodic. On the other hand, the classical time evolution
of chaotic systems is described by a non degenerate continuous spectrum,
which is capable of rapid dephasing and decay (Cornfeld,
Fomin and Sinai 1982).
Since according to Ehrenfest's theorem, classical and quantum observables
evolve alike for short times, one expects to see differences between
classical and quantum behaviour after times $T$ sufficiently long to resolve
the discreteness of the quantum spectrum.
\footnote{This is an optimist's estimate. A pessimist might estimate the
turnover time from the evolution of Gaussian wavepackets and conclude
that it goes like $\log (1/\hbar)$
because of exponential spreading (Berry and Balazs 1979).
Recent work by Tomsovic and Heller (1991)
seems to support the optimist's version.}
One now could try to develop measures that characterize the transient
chaos-like properties of quantum systems whose classical counterpart
shows chaos
\footnote{Since there is no intrinsic definition, the correspondence
to a classical system is required, with all its limitations and
pecularities as for example in the case of spin systems
(Graham and H\"ohnerbach 1984).} or focus on generic properties
of quantum systems in analogy to chaos being the generic property of
classical systems (e.g.~Eckhardt 1988a).

Alternatively, one may use the association `chaos $=$ irregularity'
and ask whether one can identify within wave theory properties
that give rise to irregularities of some kind. A  hint is provided by
wave optics: a superposition of a sufficient number of rays or waves
leads to irregular speckle patterns.
This analogy has been studied by O'Connor, Gehlen and Heller (1987),
who also found a difference between
patterns due to interference of waves with random amplitudes
and phases (as in the usual optical speckles) or due to waves with random
phases but fixed amplitudes (the case relevant for quantum
waves in billiards).
This line of thought leads to random matrix theory (Haake 1991).

For quantum systems, the equivalent of light rays is provided by the
semiclassical propagator, which expresses the transition amplitude
from one point to another in a given time interval
as a sum over all classically allowed paths connecting the two points
in that time (Miller 1974). This semiclassical propagator in the time domain
has attracted some attention recently (Tomsovi{\'c} and Heller 1991).
In most cases, however, one transforms from the time domain to the energy
domain and takes the trace which then allows one to study the spectrum of
energy eigenvalues of the system (Gutzwiller 1967, 1969, 1970, 1971, 1990).
In one degree of freedom one ends up with the WKB
method (for phase shifts, propagators, quantization rules etc.), which is
known to provide accurate and useful estimates (Berry and Mount 1972).

In many degree of freedom chaotic systems the semiclassical analysis
leads to a relation between the quantum eigenvalues and classical periodic
orbits (Gutzwiller 1990).
The analysis is not easy to carry through to
the end, with a quantization rule as simple as WKB as the final answer,
since the number of periodic orbits grows exponentially with period, thus
prohibiting a direct evaluation (Eckhardt and Aurell 1989).
To some extend this exponential proliferation
is the cause for irregular features in quantum systems,
especially in the wavefunctions (Voros 1976, Berry 1977).

However, one also has a semiclassical expression, which in principle
should be quantitatively useful when Planck's constant is sufficiently small,
the de Broglie wavelength sufficiently short. The problem is how to
tame the exponential proliferation of terms, that is, how to
overcome the `topological pressure' (Ruelle 1978, Gaspard 1992)
or the `entropy barrier' (Berry and Keating 1990).
This problem occurs in a variety of contexts also outside semiclassical
mechanics and techniques have been developed to overcome exactly
this proliferation problem (Artuso, Aurell and Cvitanovi{\'c} 1990a,b).
Below I will demonstrate
both the taming of exponential proliferation of classical orbits
and the emergence of irregular features
for the time delay in scattering. I will do so for a specific model,
scattering off three disks, that has proven extremely useful in developping and
testing a number of concepts in chaotic scattering (Eckhardt 1987,
Gaspard and Rice 1989a,b,c, Cvitanovi{\'c} and Eckhardt 1989, Eckhardt
et al 1992, Eckhardt and Russberg 1992, Tanner et al 1991).

\bigskip

\centerline{WIGNER'S TIME DELAY}

The quantity for which I shall discuss the semiclassical
expansion is Wigners time delay (Wigner 1955).
As discussed by Wigner (single scattering channel) and Smith (1960)
(many channels),it describes the phase difference between a scattered wave
and a freely propagating one (though with the same asymptotic motion
and thus with perhaps a single reflection in the interaction region,
cf. Narnhofer and Thirring 1981). It controlls the absorption in
a scattering experiment (Doron and Smilansky 1992, Doron, Smilansky and
Frenkel 1991) and is of some relevance to persistent currents in
mesoscopic rings (Akkermans et al 1991).

In terms of the $S$-matrix, the time delay $\tau(k)$ is given by
\beq
\tau = - i \hbar {\rm\  tr\ } S^\dagger {\partial S \over \partial E}
= - i \hbar {\partial \over \partial E} \ln \det S \, .
\eeq
The semiclassical expression is identical to Gutzwillers (1990)
density of states (i.e.\ the excess density of states above
a uniform background) and has been mentioned by Balian and Bloch (1974).
For the billiard problem studied here, energy and Planck's constant
only appear in the combination of the wavenumber, $k = \sqrt{2mE}/\hbar$
($m$ is the mass of the particle).
When expressed in terms of wavenumber rather than energy, the time-delay
becomes a `length'-delay, viz.
\beq
\tau(k) = \tau_0(k) + {\rm Re\ } \sum_p \sum_{r=1}^\infty L_p
{e^{i L_p k r - i \mu_p \pi r / 2} \over \sqrt{|\det(1-M_p^r)|}}  \,.
\label{semiclas}
\eeq
As usual in periodic orbit theory (Gutzwiller 1990, Eckhardt 1992),
the semiclassical expression splits into a smooth part ($\tau_0(k)$)
and a fluctuating part ($\tau_f(k)$) determined by periodic orbits.
Since we have expressed everything in terms of wavenumber $k$ rather
than energy $E$, the time delay contains the geometrical length $L_p$
of the trapped trajectory, the Maslov phase $\mu_p$ and the
linearization perpendicular to the orbit $M_p$.
The summation on $r$ accounts for multiple traversals of an orbit.
This expression is very similar to the classical trapping time distribution,
where the amplitude of a trapped periodic orbit would be $1/|\det(1-M_p^r)|$,
i.e.\ without the square root and phases
(Kadanoff and Tang 1984). This difference reflects
the one between a quantum amplitude and a classical probability.

For the two degree of freedom systems considered below, $M_p$
has eigenvalues $\Lambda_p$ and $1/\Lambda_p$ (by convention, $|\Lambda_p|>1$).
Using an expansion of Miller (1975) for the
determinant, one can express of the time delay as
a logarithmic derivative of an infinite product:
\bea
\tau_f(k) &=& {\rm Re\ } \sum_p \sum_{r=1}^\infty {{- i \over r} {\partial
\over \partial k} }
{e^{i L_p k r - i \mu_p \pi r / 2} \over \sqrt{|\det(1-M_p^r)|}}
\\
&=&  {\rm Re\ } i{\partial\over \partial k}
\sum_p \sum_{j=0}^\infty \sum_{r=1}^\infty \left( {- {1 \over r}}\right)
\left( { e^{i L_p k - i \mu_p \pi / 2}
|\Lambda_p|^{-1/2} \Lambda_p^{-j}} \right)^r
\\
&=&  {\rm Re\ } i{\partial\over \partial k}
\sum_p \sum_{j=0}^\infty \log\left( 1 -
{ e^{i L_p k - i \mu_p \pi / 2}
|\Lambda_p|^{-1/2} \Lambda_p^{-j}} \right)
\\
&=&  {\rm Re\ } i{Z'(k) \over Z(k)}
\eea
where the prime denotes a derivative with respect to wavenumber $k$ and
\beq
Z(k) =
\prod_p \prod_{j=0}^\infty \left( 1 -
{ e^{i L_p k - i \nu_p \pi / 2}
|\Lambda_p|^{-1/2} \Lambda_p^{-j}} \right)
\label{zeta_p}
\eeq
is a Selberg type zeta function (Selberg 1956, Voros 1988), named after
similar expressions in the theory of motion on surfaces of constant
negative curvature (Balazs and Voros 1986).

The technical difficulty with this expression is the exponential
proliferation of periodic orbits, as mentioned before.
For the three disk billiard, this is easily seen from the
topology of trajectories:
After leaving a disk, the particle has two choices for the
next collision. Thus the number of orbits grows
like $3\cdot 2^{n-1}$ with $n$ the number of collisions (there are
three choices for the first and two for the following ones)
(Eckhardt 1987, 1992). Taking into account an average length of paths
between collisions, this translates into an exponential proliferation
with length (Fig.~1).

\begin{figure}
\unitlength1cm
\begin{center}
\begin{picture}(6.,7.) \end{picture}\par

\begin{quote}
{\bf Fig.~1}:
Proliferation of periodic orbits with period for the three disk system
for different distance to radius separations. From top to bottom:
$d:R$ $=$ $2.5$, $2.8$, $3$, $4$, $5$ and $6$.
\end{quote}

\end{center}
\end{figure}

For practical calculations it is now useful to group terms
according to their topological length. Technically, one introduces
a counting variable $z^{n_p}$ where $n_p$ is the topological period
of the orbit and expands
\beq
Z(k) = \sum_p \sum_{r=1}^\infty L_p
{e^{i L_p k r - i \nu_p \pi r / 2} \over \sqrt{|\det(1-M_p^r)|}} z^{n_p r}
= \sum_n C_n(k) z^n \,\
\label{Selberg}
\eeq
in a power series using formulas of Plemelj (1909) and Smithies (1941)
(see also Reed and Simon 1983).
In Figure 2 the coefficients $C_n(k)$ for various distance
to radius ratios within the $A_1$ symmetry are shown.
One clearly notes the rapid decay of coefficients with increasing $n$.

\begin{figure}
\unitlength1cm
\begin{center}
\begin{picture}(6.,7.) \end{picture}\par

\begin{quote}
{\bf Fig.~2}:
Scaling of the coefficients in the cycle expanded Selberg zeta
product (\ref{Selberg}). Shown is $\log_{10} |C_n(k=0)|$ vs. $n$
for the same distance to radius ratios as in Figure~1.
\end{quote}

\end{center}
\end{figure}

As a careful analysis of the coefficients $C_n(k)$ in
(\ref{Selberg}) shows, all but the first few are combinations
of long orbits with short orbits approximating the long ones
(Cvitanovi{\'c} and Eckhardt 1989, Artuso, Aurell
and Cvitanovi{\'c} 1989a,b, Eckhardt 1992, Eckhardt and Russberg 1992).
If the index ${a^nb}$ describes the long orbit composed of
$n$ repetitions of a shorter one ${a}$ and a second
orbit $b$, then the expansion (\ref{Selberg}) will contain
contributions of the form (e.g.~for $j=0$)
\beq
e^{i L_{a^nb} k - i \mu_{a^nb} \pi / 2} |\Lambda_{a^nb}|^{-1/2}
\left( { 1 - e^{i (L_{a^{n-1}b} + L_a - L_{a^nb}) k }
e^{- i (\mu_{a^{n-1}b} + \mu_a - \mu_{ab}) \pi / 2}
\left| {\Lambda_{a^nb} \over
\Lambda_{a^{n-1}b} \Lambda_{a}}\right|^{1/2}
}\right)
\label{diff}
\eeq
The second term in the parenthesis contains only terms comparing
the long orbit with the two approximants. In hyperbolic systems,
orbits are by definition exponentially unstable. Thus to
stay near an orbit for $n$ traversals requires one to be exponentially close
to it. Thus the difference between orbits $a^nb$ and $a^{n-1}b$
is one more traversal very (exponentially) close to $a$.
Thus the differences in action are exponentially close to zero,
the ratios in instabilities close to $1$.
If now the Maslov phases (a discrete quantity) fit as well
(one can assure this by a proper choice of symbolic organization of
orbits, Eckhardt and Wintgen 1991, Eckhardt 1992), all that enters
in (\ref{diff}) are the deviations from one, i.e.~an exponentially small
quantity. This is enough to turn the expanded
product (\ref{Selberg}) convergent for real energies.

This is a first example of order among all this chaos: long periodic
orbits are not random numbers but are strictly correlated with
shorter orbits, so that one can actually estimate the contributions
from long orbits rather accurately and obtain convergent results
for the density of states and related quantities.

For large $n$ the decay is only exponential rather than superexponential, in
contrast to the classical Selberg zeta function (Cvitanovi{\'c} and
Eckhardt 1991). Techniques have been developed to improve
this further to recover faster than exponential convergence
(Eckhardt and Russberg 1992). A new semiclassical Selberg type
product, similar to the classical one, is currently under investigation.
Preliminary test indicate improved convergence.
The calculations shown here were performed with the expansion
(\ref{Selberg}) directly, using orbits of period up to 13.

One test of the semiclassical expression is to compare zeros of
(\ref{Selberg}) with poles of the $S$-matrix. The result (Cvitanovi{\'c} and
Eckhardt 1989, Eckhardt et al 1992) is that the differences
between exact and semiclassical resonances decrease with increasing
wavenumber, i.e. when approaching the semiclassical limit.
This is in spirit with the semiclassical approximation and confirms
that Gutzwillers (1990) analysis is a bona fide semiclassical theory.

A plot of the time delay for a larger interval in $k$ shows
large fluctuations, and lots of irregular features (Fig.~3).
They come about because every resonance of the $S$-matrix contributes
a Lorentzian to the time delay.
If the positions of the resonances are $k_i = s_i - i \gamma_i$,
then
\beq
\tau(k) = \sum_i {\gamma_i \over (k-s_i)^2 + \gamma_i^2} \, .
\label{reson}
\eeq
Cross sections differ from the above expression in that the
weights of the Lorentzians are not unity (Ericson 1960,
Brink and Stephen 1963) and also show fluctuations.
Signals of this kind were extensively studied in nuclear physics
(Ericson 1960, Brink and Stephen 1963) in connection with the compound
nucleus. It was shown that a correlation function contains information
on the average widths of resonances and thus on the lifetime of
the compound nucleus. A semiclassical interpretation has been
attempted by Bl\"umel and Smilansky (1988), starting directly from
the $S$-matrix. In this limit, however, the same information
should also be contained in the time delay and its
correlation function (for a discussion of
deviations in extreme quantum cases, see Lewenkopf and Weidenm\"uller 1991).

\begin{figure}
\unitlength1cm
\begin{center}
\begin{picture}(6.,12.) \end{picture}\par

\begin{quote}
{\bf Fig.~3}:
Time delay for scattering off the three disk system at $d:R=2.5$.
The lower frame shows a magnification and comparison between
different maximal periods of orbits included.
\end{quote}

\end{center}
\end{figure}

\bigskip

\centerline{CORRELATION FUNCTIONS}

Consider the fluctuating part of the time delay (after subtracting
the mean), more precisely a segment $\tau_s(k) = \tau_f(k) w(k)$,
projected out
with a window function $w(k)$ (e.g.~a box of width $\Delta k$ centered
around $k_0$ or a Lorentzian). The autocorrelation function of this
segment is then given by the integral
\beq
C(\kappa) = \int_{\Delta k} dk\  \tau_s(k-\kappa/2)\, \tau_s(k+\kappa/2) \ .
\label{C(k)}
\eeq
Its Fourier transform $K(\lambda)$ is the structure function,
\beq
K(\lambda) = \int d\kappa \ C(\kappa)\  \cos(\kappa \lambda)\, .
\eeq

Since the time delay is a superposition of Lorentzians, the
correlation function for small $\kappa$ probes their widths.
In a simplified analysis,
assume that the window $w(k)$ selects a finite number
of resonances in each $\tau(k)$. Then
\beq
C(\kappa) = \sum_i \sum_j \int dk\
{\gamma_i \over (k-\kappa/2-s_i)^2 + \gamma_i^2} \
{\gamma_j \over (k+\kappa/2-s_j)^2 + \gamma_j^2} \
\eeq
which, in the diagonal approximation, becomes
\beq
C(\kappa) = \sum_i
{2 \gamma_i \over (\kappa-s_i)^2 + 4 \gamma_i^2}
\approx
{2 \gamma \over \kappa^2 + 4 \gamma^2} \,.
\label{corr}
\eeq
The last form is an approximation by a single Lorentzian of
effective width $2 \gamma$.

\begin{figure}
\unitlength1cm
\begin{center}
\begin{picture}(6.,7.) \end{picture}\par

\begin{quote}
{\bf Fig.~4}:
Autocorrelation function of the time delay for $d:R=2.5$.
It was computed for intervals of $\Delta \kappa=50$ starting
at the $k$-values indicated in the figure. For comparison, the bold
line shows a Lorentzian of width $\gamma=0.15$ (Eq.~\ref{corr}).
\end{quote}

\end{center}
\end{figure}

Comparison with the computed
autocorrelation function (Figure~4)
shows that a Lorentzian describes the correlation function only
for very small $k$. For larger ones, $C(\kappa)$ even becomes negative.
A similar discrepancy was noted by  Wardlaw and Jaworski (1989)
in their analysis of the phase shift for a particle scattered
off a leaky surface of constant negative curvature (a model
introduced by Gutzwiller 1983). In a later study,
Shushin and Wardlaw (1992) could actually
improve on the form of $C(\kappa)$ by
adding correlations between the positions of
resonances, predicting a correlation function of the form
\beq
C_{SW} (\kappa)\approx  \gamma^2
{\gamma^2-\kappa^2 \over (\kappa^2 + \gamma^2)^2} \,.
\eeq
This form does not fit the data any better, presumably because
the underlying ensemble of resonances is different from the
assumed $GUE$ ensemble (see e.g.~Haake 1991).

The large $\kappa$ behaviour of the correlation function is
difficult to access from the resonance representation (\ref{reson}).
This behaviour on the other hand is reflected in the
small $\lambda$ behaviour of the structure function.
Using the semiclassical representation (\ref{semiclas}) for the time delay,
the structure function for a segment $\tau_s(k)$ becomes
\bea
K(\lambda) &=& \left| {\sum_p \sum_{r=1}^\infty
{L_p e^{- i \nu_p \pi r / 2} \over \sqrt{|\det(1-M_p^r)|}}
\tilde{w}(\lambda-rL_p)
}\right|^2
\\
&=& \sum_p\sum_{r=1}^\infty
{L_p^2 \over |\det(1-M_p^r)|} \tilde{w}^2(\lambda-rL_p)  +
\sum_P \sum_{P'} A_{P} A^*_{P'} \tilde{w}(\lambda-L_P) \tilde{w}(\lambda-L_p)
\label{sft}
\, ,
\eea
where $\tilde{w}$ is the Fourier transform of the window function;
$P$ and $P'$ denote periodic orbits (perhaps multiple traversals
of shorter ones) with weights $A_P$, $A_{P'}$ and of length $L_P$
and $L_{P'}$, respectively.
If the window is sufficiently wide, then the Fourier transform
will be narrow and no terms will overlap when taking the
absolute value squared. Therefore, only the diagonal contributions in
(\ref{sft}) survive and the structure function will show individual periodic
orbits for small $\lambda$. For very large $\lambda$,
interferences between different periodic orbits
will eventually turn the sum into the exponential decay required
by the Lorentzian result for the small $\tau$ behaviour of the correlation
function.

\begin{figure}
\unitlength1cm
\begin{center}
\begin{picture}(6.,7.) \end{picture}\par

\begin{quote}
{\bf Fig.~5}:
Structure function for the time delay shown in Figure~3.
The inset shows the structure function over a larger range
on a logarithmic scale. Clearly visible is the exponential
fall off related to the Lorentzian of the autocorrelation function.
\end{quote}

\end{center}
\end{figure}

\bigskip
\centerline{RELATION TO BOUNDED SYSTEM}

As the disks are moved close enough to touch, they enclose a region
of the shape of a tipped triangle. Since this is a bounded system,
it has a discrete spectrum. The Wigner time delay
then turns into the density of states of the bounded system and
the resonances become infinitely sharp. Semiclassical quantization
attempts for this system have been reported elsewhere (Tanner et al 1991).
Here I would like to describe the relation between the
correlation function and structure function for the time delay
as found above to the semiclassical theory of two point correlation
functions by Berry (1985).

The small $\lambda$ behaviour of the structure function with the periodic
orbits is in agreement with Berry's semiclassical analysis and
numerical results (e.g.~Wintgen 1987). The large $\lambda$
exponential decay is consistent with it, since the decay rate
is related to the width of the resonances; in a bounded system, the
widths goes to zero and $K(\lambda)$ settles to a constant
(Levandier et al 1986, Pique et al 1987).
What seems to be missing is the intermediate regime, where
in a bounded system $K(\lambda)$ should increase linearly.

For $\lambda$ large enough so that the periodic orbits
overlap, but not too large so that the cancellations take over,
the correlation function near $\lambda$ sums over all
orbits with length near $\lambda$,
\beq
K(\lambda) \approx  \sum_{L_p {\rm\  near\ }\lambda}
{L_p^2 \over |\det(1-M_p^r)|} \approx  \lambda e^{-\Gamma \lambda}\,.
\eeq
The approximate form is valid for intermediate $\lambda$ and contains the
classical escape rate $\Gamma$
(Kadanoff and Tang, 1984, Cvitanovi{\'c} and Eckhardt 1991),
up to a factor $L$. In the case of bounded systems, $\Gamma=0$ and $K(\lambda)$
increases linearly with $\lambda$. For open systems, it
attains a maximum near $\lambda = 1/\Gamma$. For the three disk
system at $d:R=2.5:1$, $\Gamma\approx 0.7$, so that it is impossible
to detect the intermediate classical region. However, the
exponent $\Gamma$ clearly does not describe the exponential decay for
large $\lambda$, as speculated by Bl\"umel and Smilansky (1988).
Incidently, notice that in this case the quantum decay is {\em slower}
than the classical decay. Thus the interferences between off-diagonal
terms in (\ref{sft}) act to increase $C(\lambda)$ compared to the
classical value and not to decrease it as in the case
of the bounded system (Berry 1985).

\bigskip

\centerline{CONCLUSIONS}

Investigations in `quantum chaos' and in particular on
Gutzwiller's periodic orbit formula for the density of states
have helped to uncover a
considerable amount of order among irregular looking quantum data.
The main tool is the Fourier transform which facilitates the transition
from the energy domain to the time domain, where the dynamics is
dominated by correlations; they give rise to longer range
correlations in energy. I have demonstrated these ideas using
as an example a simple scattering system, where
one has good control over both  classical and quantum dynamics and where
it is possible to sum the semiclassical expressions, despite the
exponential proliferation of orbits.  The results confirm
that the cycle expansion of Selberg
products (\ref{Selberg}) provides a convenient and accurate
method for the calculation of periodic orbit sums in hyperbolic systems.
It has allowed for a accurate test of Gutzwiller' semiclassical theory.

A number of problems still remain. Among them are the extension of
periodic orbit summation techniques to non-hyperbolic cases,
the summation of expressions for cross sections and the extension to more
degrees of freedom than just two (perhaps even field theories).
I am optimistic that eventually,
a semiclassical theory for chaotic systems, perhaps as useful and
reliable as the well-known WKB methods, will emerge.

\bigskip

\centerline{ACKNOWLEDGEMENT}

The above presentation has been influenced very much by discussions
with Shmuel Fishman.


\newcommand{\PR}[1]{{\em Phys.\ Rep.}\/ {\bf #1}}
\newcommand{\PRL}[1]{{\em Phys.\ Rev.\ Lett.}\/ {\bf #1}}
\newcommand{\PRA}[1]{{\em Phys.\ Rev.\ A}\/ {\bf #1}}
\newcommand{\PRD}[1]{{\em Phys.\ Rev.\ D}\/ {\bf #1}}
\newcommand{\JPA}[1]{{\em J.\ Phys.\ A}\/ {\bf #1}}
\newcommand{\JPB}[1]{{\em J.\ Phys.\ B}\/ {\bf #1}}
\newcommand{\JCP}[1]{{\em J.\ Chem.\ Phys.}\/ {\bf #1}}
\newcommand{\JPC}[1]{{\em J.\ Phys.\ Chem.}\/ {\bf #1}}
\newcommand{\JMP}[1]{{\em J.\ Math.\ Phys.}\/ {\bf #1}}
\newcommand{\AP}[1]{{\em Ann.\ Phys.}\/ {\bf #1}}
\newcommand{\PL}[1]{{\em Phys.\ Lett.}\/ {\bf #1}}
\newcommand{\PLA}[1]{{\em Phys.\ Lett.\ A}\/ {\bf #1}}
\newcommand{\PLB}[1]{{\em Phys.\ Lett.\ B}\/ {\bf #1}}
\newcommand{\PD}[1]{{\em Physica D}\/ {\bf #1}}
\newcommand{\NPB}[1]{{\em Nucl.\ Phys.\ B}\/ {\bf #1}}
\newcommand{\INCB}[1]{{\em Il Nuov.\ Cim.\ B}\/ {\bf #1}}
\newcommand{\JETP}[1]{{\em Sov.\ Phys.\ JETP}\/ {\bf #1}}
\newcommand{\JETPL}[1]{{\em JETP Lett.\ }\/ {\bf #1}}
\newcommand{\RMS}[1]{{\em Russ.\ Math.\ Surv.}\/ {\bf #1}}
\newcommand{\USSR}[1]{{\em Math.\ USSR.\ Sb.}\/ {\bf #1}}
\newcommand{\PST}[1]{{\em Phys.\ Scripta T}\/ {\bf #1}}
\newcommand{\CM}[1]{{\em Cont.\ Math.}\/ {\bf #1}}
\newcommand{\JMPA}[1]{{\em J.\ Math.\ Pure Appl.}\/ {\bf #1}}
\newcommand{\RMP}[1]{{\em Rev.\ Mod.\ Phys.}\/ {\bf #1}}

\bigskip

\centerline{REFERENCES}

\frenchspacing
\begin{tabbing}
Akkermans, E., Auerbach A., Avron, J.E. and Shapiro. B (1990) \PRL{66}, 76\\
Artuso, R., Aurell, E. and Cvitanovi{\a'c}, P. (1990a)
             {\em Nonlinearity} {\bf 3}, 325\\
----- (1990b) {\em Nonlinearity} {\bf 3}, 361 \\
%
%
%
Balazs, N. and Voros, A. (1986) \PR{143}, 109\\
Balian, R. and Bloch, C. (1974) \AP{85} 514\\
Berg{\a'e}, P., Pomeau, Y. and Vidal, Ch. (1984) {\em L'ordre dans le chaos},
	Hermann, Paris\\
Berry, M.V. (1977) \JPA{12}, 2083\\
----- (1985) {\em Proc. R. Soc. (London)} {\bf A 400}, 229\\
Berry, M.V. and Balazs, N. (1979) \JPA{12}, 625\\
Berry, M.V. and Keating, J.P. (1990) \JPA{23}, 4839\\
Berry, M.V. and Mount, K.E. (1972) {\em Rep. Prog. Phys.} {\bf 35}, 315\\
Bl\"umel, R. and Smilansky, U. (1988) \PRL{60}, 477\\
%
%
Brink, D.M. and Stephen, R.O. (1963) {\em Phys. Lett.} {\bf 5}, 77\\
Campbell, D. and Rose, H. (1983) (eds) {\em Order in Chaos}, \PD{7}\\
Cornfeld, I.P., Fomin, S.V. and Sinai, Ya.G. (1982) {\em Ergodic Theory},
	Springer, Berlin\\
Cvitanovi{\a'c}, P. and Eckhardt, B. (1989) \PRL{63}, 823 \\
----- (1991) \JPA{24}, L237 \\
%
%
%
%
%
Doron, E. and Smilansky, U. (1992) \PRL{68}, 1255\\
Doron, E., Smilansky, U. and A. Frenkel (1990) \PRL{65}, 3072\\
Eckhardt, B. (1987) \JPA{20}, 5971 \\
----- (1988) {\em Phys. Rep.} {\bf 163}, 205\\
%
%
----- (1992) {\em Periodic orbit theory}, to appear in
                  {\em Proceedings of the International School of} \\
\ \ \ \ \ \ \  {\em Physics ``Enrico Fermi'' 1991, Course CXIX}
              {\em ``Quantum Chaos''}\/, \\
\ \ \ \ \ \ \  G.~Casati, I.~Guarneri,
               and U.~Smilansky (eds).\\
Eckhardt, B. and Aurell, E. (1989) {\em Europhys. Lett.} {\bf 9} 509\\
Eckhardt, B. and Russberg, G. (1992) {\em Resummation of classical and
semiclassical}\\
\ \ \ \ \ \ \   {\em periodic orbit formulas}, preprint\\
Eckhardt, B., Cvitanovi{\'c}, P., Rosenqvist, P. E., Russberg, G. and
	Scherer, P. (1992b)\\
\ \ \ \ \ \ \   {\em Pinball scattering}\/,
                  to appear in {\em Quantum Chaos}\/, \\
\ \ \ \ \ \ \   G.~Casati, B.~V.~Chirikov
                  (eds), Cambridge University Press\\
Eckhardt, B. and Wintgen, D. (1991) \JPA{23}, 4335\\
%
%
Ericson, T. (1960) \PRL{5}, 430\\
Gaspard, P. (1992) {\em Scattering and resonances: Classical and quantum
		dynamics},  to appear in\\
\ \ \ \ \ \ \    {\em Proceedings of the International School of Physics
                  ``Enrico Fermi'' 1991, Course CXIX}\\
\ \ \ \ \ \ \   {\em   ``Quantum Chaos''}\/,
       		G.~Casati, I.~Guarneri, and U.~Smilansky (eds).\\
Gaspard, P. and  Rice, S.A. (1989a) \JCP{90}, 2225 \\
----- (1989b)       \JCP{20}, 2242\\
----- (1989c)       \JCP{20}, 2255\\
Graham, R. and H\"ohnerbach, M. (1984) {\em Z. Phys B.} {\bf 57}, 233\\
Gutzwiller, M.C. (1967) \JMP{8}, 1979 \\
----- (1969)        \JMP{10}, 1004\\
----- (1970)        \JMP{11}, 1791\\
----- (1971)        \JMP{12}, 343\\
----- (1983)        {\em Physica D} {\bf 7}, 341\\
----- (1990) {\em Chaos in Classical and Quantum Mechanics}\/,
 		Springer, New York\\
Haake, F. (1991) {\em Quantum signatures of chaos}, Springer, Berlin\\
%
%
Kadanoff, L.P. and Tang, C. (1984) {\em Proc. Natl. Acad. Sci. USA}
	{\bf 81}, 1276\\
%

%
Levandier, L., Lombardi, M., Jost, R. and Pique J.P. (1986) \PRL{56}, 2449\\
Lewenkopf, C.H. and Weidenm\"uller, H.A. (1991)
		{\em Ann. Phys. (NY)} {\bf 212} 53\\
Miller, W.H. (1974) {\em Adv. Chem. Phys.} {\bf 25}, 69 \\
----- (1975) \JCP{63}, 996\\
%
%
Narnhofer, H. and Thirring, W. (1981) \PRA{23}, 1688\\
O'Connor, P., Gehlen, J. and Heller, E.J. (1987) \PRL{58}, 1296\\
Pique, J.P., Chen, Y., Field, R.W. and Kinsey, J.L. (1987) \PRL{58}, 475\\
Plemelj, J. (1909) {\em Monat.\ Math.\ Phys.}\/ {\bf 15}, 93\\
Reed, M. and Simon, B. (1978) {\em Methods of Modern Mathematical
                  Physics}\/ vol 4\\
\ \ \ \ \ \ \   {\em Analysis of Operators}, Academic Press, New York\\
Ruelle, D. (1978) {\em Thermodynamic Formalism},
                  Addison-Wesley (Reading 1978)\\
Schuster, H.G. (1988) {\em Deterministic Chaos}, VCH Publishers, Weinheim 1988\\
Selberg, A. (1956) {\em J.\ Ind.\ Math.\ Soc.}\/ {\bf 20}\/, 47\\
Shushin, A. and Wardlaw, D.M. (1992) \JPA{25}, 1503\\
Smith, F.T. (1960) {\em Phys. Rev.} {\bf 118}, 349\\
Smithies, F. (1941) {\em Duke Math.\ J.}\/ {\bf 8} 107\\
Tanner, G., Scherer, P., Bogomolny, E.B., Eckhardt, B. and Wintgen, D. (1991)\\
\ \ \ \ \ \ \ \   \PRL{67}, 2410\\
Tomsovic, S. and Heller, E.J. (1991) \PRL{67}, 664\\
Voros, A. (1977) {\em Ann. Inst. H. Poincar{\a'e} A} {\bf 26}, 343\\
----- (1988) \JPA{21}, 685\\
Wardlaw, D.M. and Jaworski, W. (1989) \JPA{22}, 3561\\
Wintgen, D. (1987) \PRL{58}, 1589\\
Wigner, E.P. (1955) {\em Phys. Rev.} {\bf 98}, 145\\
\end{tabbing}

\end{document}